\pdfoutput=1

\documentclass[aps,prstab,reprint,groupedaddress,showpacs,floatfix,a4paper]{revtex4-1}

\usepackage{graphicx}
  \setkeys{Gin}{width=\linewidth}
\usepackage{amsmath,amsfonts,amssymb}
\usepackage{amsbsy}

\usepackage{CJKutf8}

\usepackage{epstopdf}%This line makes .eps figures into .pdf - please comment out if not required.

\begin{document}

%\preprint{}

\title{Landau quantization in buckled monolayer GaAs}
%\title{Landau quantization in monolayer GaAs}
%Selection rules of optical absorption spectra in graphene nanoribbons


\author{Hsien-Ching Chung}
\email[E-mail: ]{hsienching.chung@gmail.com}
\homepage[\\ Homepage: ]{https://sites.google.com/site/hsienchingchung/home}
%\thanks{}
\affiliation{Department of Physics, National Kaohsiung Normal University, Kaohsiung 824, Taiwan}
%\author{Ching-Hong Ho}
%\email[]{hohohosho@gmail.com}
%\affiliation{Center for General Education, Tainan University of Technology, Tainan 701, Taiwan}
%\author{Cheng-Peng Chang}
%\email[]{t00252@mail.tut.edu.tw}
%\affiliation{Center for General Education, Tainan University of Technology, Tainan 701, Taiwan}
\author{Chun-Nan Chen}
%\email[]{quantum@mail.tku.edu.tw \& ccn1114@kimo.com}
\affiliation{Quantum Engineering Laboratory, Department of Physics, Tamkang University, Tamsui, New Taipei 25137, Taiwan}
\author{Chih-Wei Chiu}
\email[E-mail: ]{giorgio@fonran.com.tw}
\affiliation{Department of Physics, National Kaohsiung Normal University, Kaohsiung 824, Taiwan}
\author{Ming-Fa Lin}
\email[E-mail: ]{mflin@mail.ncku.edu.tw}
\affiliation{Department of Physics, National Cheng Kung University, Tainan 70101, Taiwan}


\date{\today}

\begin{abstract}
Magneto-electronic properties of buckled monolayer GaAs is studied by the developed generalized tight-binding model, considering the buckled structure, multi-orbital chemical bondings, spin-orbit coupling, electric field, and magnetic field simultaneously.
Three group of spin-polarized Landau levels (LLs) near the Fermi level are induced by the magnetic quantization, whose initial energies, LL degeneracy, energy spacings, magnetic-field-dependence, and spin polarization are investigated.
The Landau state probabilities describing the oscillation patterns, localization centers, and node regularities of the dominated/minor orbitals are analyzed, and their energy-dependent variations are discussed.
The given density of states directly reflects the main features of the LL energy spectra in the
structure, height, number, and frequency of the spin-polarized LL peaks.
The electric field causes the monotonous/nonmonotonous LL energy dispersions, LL crossing, gap modulation, phase transition and spin splitting enhancement.
The complex gap modulations and phase transitions based on the competition between magnetic and electric fields are explored in detail by the phase diagram.
The field-controlled gap modulations and phase transitions are helpful in designing the top-gated and phase-change electronic devices.
These predicted magneto-electronic properties could be verified by scanning tunneling spectroscopy measurements.
%Furthermore, this work can serve as a model study for understanding magnetic quantizations of other group III-V 2D materials.
\end{abstract}

%\pacs{78.66.Tr, 78.67.Lt, 71.15.Ap}
%\keywords{}

\maketitle

\section{Introduction}

%%-----------------------------------
%1st SUBSECTION
%
%emergent 2D materials, covering group-IV, group-V, group III-V compounds, TMD, and few-layer black phosphorene
%
%Such 2D materials/layered systems/layered materials are expected to have the rich and essential properties, being sensitive to the lattice symmetry, stacking configuration, layer number, orbital hybridization, SOC, and external electric and magnetic fields.
%This work is focused on the magnetic quantization of monolayer GaAs using the generalized tight-binding model.
%Comparisons to other 2D systems are also made.
%%-----------------------------------

Over the past decade, graphene~\cite{Science306(2004)666K.S.Novoselov, Proc.Natl.Acad.Sci.U.S.A.102(2005)10451K.S.Novoselov} has successfully brought scientists into the world of two-dimensional (2D) materials based on its incredible intrinsic properties, such as high carrier mobility at room temperature ($>$ 200000 cm$^2$/Vs)~\cite{Science312(2006)1191C.Berger, SolidStateCommun.146(2008)351K.I.Bolotin, Phys.Rev.Lett.100(2008)016602S.V.Morozov}, superior thermoconductivity (3000--5000 W/mK)~\cite{Phys.Rev.Lett.100(2008)016602S.V.Morozov, NanoLett.8(2008)902A.A.Balandin}, high transparency for incident light over a wide range of wavelength (97.7 $\%$)~\cite{Science320(2008)1308R.R.Nair, Nat.Nanotechnol.5(2010)574S.Bae}, extremely large modulus ($\sim$1 TPa) and tensile strength ($\sim$100 GPa)~\cite{Science321(2008)385C.Lee}.
Few-layer graphene are observed to have diverse magnetic quantizations, e.g., the Landau levels (LLs) with $\sqrt{B_z}$-dependence in monolayer graphene featuring massless Dirac fermion~\cite{Phys.Rev.Lett.102(2009)176804G.Li, Science324(2009)924D.L.Miller, Nature467(2010)185Y.J.Song, Phys.Rev.Lett.106(2011)126802A.Luican, Phys.Rev.B92(2015)165420W.X.Wang}, the LLs with linear $B_z$-dependence in AB-stacked bilayer graphene featuring massive Dirac fermion~\cite{Phys.Rev.Lett.100(2008)087403E.A.Henriksen, Nat.Phys.7(2011)649G.M.Rutter, Phys.Rev.B93(2016)125422L.J.Yin}, as well as the coexistence of square-root and linear $B_z$-dependent LLs in graphene of trilayer ABA stacking~\cite{Phys.Rev.B91(2015)115405L.J.Yin}, where $B_z$ is the strength of magnetic field.
Although interest in graphene is still high, it is also conspicuous that graphene has its limitation.
For instance, in contrast to conventional semiconductors, the lack of a significant band gap limits its applicability in electronics where high transistor on/off ratios are vital~\cite{NanoLett.10(2010)715F.Xia}.
This obstacle triggers researches on emergent 2D materials~\cite{ACSNano9(2015)11509G.R.Bhimanapati, Nanoscale7(2015)8261F.Schwierz, J.Phys.D-Appl.Phys.50(2017)053004Y.B.Zhang}, covering group-IV~\cite{Phys.Rev.B87(2013)035438L.Matthes, 2DMater.2(2015)035012W.Amamou, 2DMater.3(2016)025034H.Liu}, group-V~\cite{ACSNano8(2014)4033H.Liu, Nat.Nanotechnol.9(2014)372L.Li, 2DMater.2(2015)011002J.O.Island}, group III-V compounds~\cite{Phys.Rev.B80(2009)155453H.Sahin, Phys.Rev.B87(2013)165415H.L.Zhuang, NanoLett.14(2014)2505F.C.Chuang}, and transition-metal dichalcogenides (TMDs)~\cite{Nat.Nanotechnol.7(2012)699Q.H.Wang, J.Phys.Chem.C116(2012)8983C.Ataca, ACSNano8(2014)1102D.Jariwala, 2DMater.2(2015)015006G.Plechinger}.
Such 2D layered materials are expected to have the rich and essential properties, being sensitive to the lattice symmetry, stacking configuration, layer number, orbital hybridization, spin-orbit coupling (SOC), as well as external electric and magnetic fields.
This work is focused on the magnetic quantization of monolayer GaAs using the generalized tight-binding model.
%Comparisons to other 2D systems are also made.
Comparisons to graphene are also made.

%%-----------------------------------
%
%2nd SUBSECTION
%
%%-----------------------------------
%
%group-IV (C, Si, Ge, Sn) / hexagonal symmetric structure / [bucking structure, grows from Si to Sn ]
%
%group-IV EXP examples
%
%graphene has plane hexagonal structure, and the others have buckled structure and SOC
%buckled structure and SOC grows as the atomic mass increases
%
%SOC and the hybridization of multi-orbitals dominate the low-energy electronic structure
%
%graphene is a zero-gap semiconductor, while the others have gaps
%
%group-IV exhibit rich magnetic quantization
%
%state energy relation of LL:
%sqrt B dependence, monolayer graphene
%linear B dependence, bilayer AB-stacked graphene
%ABA trilayer graphene possesses monolayer- and bilayer-like LL spectra and dependance
%
%As for layered GaAs compound, EXP thin film example
%
%According to first-principles calculations, monolayer GaAs possesses buckled hexagonal structure, multi-orbital chemical bonding, and the significant SOC, leading to the rich electronic properties.
%This system will exhibit diverse magnetic quantization in the presence/absence of electric fields.
%%-----------------------------------

Group-IV monoelemental 2D honeycomb materials beyond graphene, such as silicene, germanene, and stanene, have been proposed to possess a band gap owing to SOC~\cite{Phys.Rev.Lett.107(2011)076802C.C.Liu, Phys.Rev.B84(2011)195430C.C.Liu, Phys.Rev.Lett.111(2013)136804Y.Xu}.
%gapped Dirac materials
Recently, few-layer silicene, germanene, and stanene have been synthesized on various substrates: silicene on Ag(111)~\cite{Phys.Rev.Lett.108(2012)155501P.Vogt, Phys.Rev.Lett.109(2012)056804L.Chen, Nat.Nanotechnol.10(2015)227L.Tao, 2DMater.3(2016)031011P.DePadova}, Ir(111)~\cite{NanoLett.13(2013)685L.Meng}, and $\mathrm{ZrB}_2$(0001)~\cite{Phys.Rev.Lett.108(2012)245501A.Fleurence}; germanene on Pt(111)~\cite{Adv.Mater.26(2014)4820L.F.Li}, Al(111)~\cite{NanoLett.15(2015)2510M.Derivaz}, and Au(111)~\cite{NewJ.Phys.16(2014)095002M.E.Davila}; stanene on $\mathrm{Bi}_2\mathrm{Te}_3(111)$~\cite{Nat.Mater.14(2015)1020F.F.Zhu}.
Silicene, germanene, and stanene having buckled structure with SOC, which grow as the atomic mass increases, are much different from the planar hexagonal graphene without SOC.
Their low-energy electronic structures are dominated by the SOC and the hybridization of multi-orbitals.
The group-IV materials with heavy atomic masses have broad buckled angle and strong SOC, leading to a large gap, e.g., the gap of germanene (stanene) is comparable to (larger than) the thermal energy at room temperature~\cite{Phys.Rev.B84(2011)195430C.C.Liu, J.Phys.Condens.Matter27(2015)443002A.Acun}.
Moreover, magnetic quantizations with various magnetic-field-dependent LLs and monotonous/nonmonotonous electric-field-dependent LLs with subband crossing/anticrossing are predicted~\cite{Phys.Rev.B88(2013)085434C.J.Tabert, Phys.Rev.B90(2014)205417V.Y.Tsaran, RSCAdv.5(2015)83350S.A.Yang, Phys.Rev.B94(2016)045410S.C.Chen, Phys.Rev.B94(2016)205427J.Y.Wu}.
However, the strong interactions between silicene (germanene, or stanene) and substrate deform the buckled structure and mix the electronic states of silicene and substrate near the Fermi level, making the modification of low-energy electronic properties.
Recent experiments on tunneling spectra of silicene closing to the liquid-helium temperature have evidenced the disappearance of LL sequences based on the instability from the dangling bonds of the $sp^3$-hybridized atoms~\cite{Phys.Rev.Lett.110(2013)076801C.L.Lin}.
%These instability results from the dangling bonds of the $sp^3$-hybridized atoms.
%resulting in the disappearance of LL sequences in the tunneling spectra of silicene under a magnetic field~\cite{Phys.Rev.Lett.110(2013)076801C.L.Lin}.

Apart from 2D materials of group-IV elements, the binary compounds of group III-V elements have also been proposed as honeycomb lattices with large gaps~\cite{Phys.Rev.B80(2009)155453H.Sahin, Phys.Rev.B87(2013)165415H.L.Zhuang, NanoLett.14(2014)2505F.C.Chuang}.
Although the group III-V elemental 2D materials of buckled structure with mixed $sp^3$--$sp^2$ bonding are more stable compared to those of planar ones with $sp^2$ bonding~\cite{Phys.Rev.B80(2009)155453H.Sahin}, the dangling-bond-induced instability remains.
A promising route is to saturate the dangling bonds by halogen atoms, which has been used in graphene~\cite{Small6(2010)2877R.R.Nair, NanoLett.10(2010)3001J.T.Robinson, ACSNano5(2011)1042K.J.Jeon}.
First-principles calculations indicate that iodinated germanene (GeI)~\cite{Phys.Rev.B89(2014)115429C.Si} and fluorinated stanene (SnF)~\cite{Phys.Rev.Lett.111(2013)136804Y.Xu} are free from dangling bonds and interact weakly with substrates.
Their gaps are about $0.3$ eV at the $\Gamma$ point, considerably larger than the values of the unpassivated 2D systems.
Bulk GaAs is one of the famous group III-V elemental binary compounds, being widely used in the manufacture of electronic and optical devices due to its direct band gap (silicon is indirect gap) and high mobility (than silicon)~\cite{Solid-StateElectron.11(1968)599S.M.Sze, IEEETrans.ElectronDevices29(1982)292N.D.Arora}.
%Bulk GaAs is one of the famous group III-V elemental binary compounds, being widely used in the manufacture of electronic and optical devices [REF] due to its direct band gap (silicon is indirect gap)[] and high mobility (than silicon)~\cite{Solid-StateElectron.11(1968)599S.M.Sze, IEEETrans.ElectronDevices29(1982)292N.D.Arora}.
%As for layered GaAs compound, [EXP thin film example]
According to first-principles calculations~\cite{Sci.Rep.5(2015)8441M.W.Zhao}, monolayer GaAs possesses buckled hexagonal structure, multi-orbital chemical bonding, and significant SOC, leading to rich electronic properties.
This system will exhibit diverse magnetic quantization in the presence/absence of electric fields.

%Electrical Characteristics of GaAs Nanowire-Based MESFETs on Flexible Plastics
%[IEEE Trans. Electron Devices 58 (2011) 1096, C. Yoon]
%Field-effect mobility of our representative MESFET is 3104.6 cm2/V.s in the flat state.
%Thickness? nm?

%Electric field induced insulator to metal transition in a buckled GaAs monolayer
%[RSC Adv. 6 (2016) 52920, B.P. Bahuguna]

%-----------------------------------

%3rd SUBSECTION

%-----------------------------------

The generalized tight-binding model built from the subenvelope functions on the layered-dependent distinct sublattices is developed to study the electronic properties under uniform/non-uniform external electric and magnetic fields.
The geometric structures, multi-orbital hybridizations, SOC, and external fields are included in the calculation, simultaneously.
The quantized energy spectra and wave functions can be efficiently computed by the method of exact diagonalization even for a rather large Hamiltonian with complex matrix elements.
This model has been widely adopted to make systematic studies on multi-dimensional carbon-based materials and hybrid systems, ranging from three-dimensional (3D) graphites~\cite{Phys.Rev.71(1947)622P.R.Wallace, Carbon43(2005)1424C.P.Chang, RSCAdv.5(2015)53736R.B.Chen}, 2D graphenes~\cite{Phys.Rev.71(1947)622P.R.Wallace, PhysicaE40(2008)1722J.H.Ho, Phys.Rev.B77(2008)085426Y.H.Lai, Phys.Chem.Chem.Phys.17(2015)26008C.Y.Lin, ACSNano9(2015)8967H.C.Wu, Nat.Commun.8(2017)14453H.C.Wu}, 1D graphene nanoribbons (GNRs)~\cite{J.Phys.Soc.Jpn.65(1996)1920M.Fujita, Phys.Rev.B59(1999)8271K.Wakabayashi, PhysicaE42(2010)711H.C.Chung, J.Phys.Soc.Jpn.80(2011)044602H.C.Chung, Phys.Chem.Chem.Phys.18(2016)7573H.C.Chung, Carbon109(2016)883H.C.Chung}, carbon nanotubes (CNTs)~\cite{Phys.Rev.B46(1992)1804R.Saito, Phys.Rev.B50(1994)17744M.F.Lin, Phys.Rev.B52(1995)8423M.F.Lin, Phys.Rev.Lett.78(1997)1932C.L.Kane, Phys.Rev.B62(2000)16092S.Roche, Phys.Rev.B67(2003)045405F.L.Shyu}, graphene nanoflake~\cite{Carbon118(2017)78K.Szalowski} and graphene-related hybrids~\cite{doi:10.1021/acsnano.7b02494}.
%, as well as CNT-GNR~\cite{Nanotechnol.19(2008)105703T.S.Li} and GNR-graphene~\cite{Synth.Met.161(2011)489C.H.Lee, Diam.Relat.Mat.20(2011)1026C.H.Lee} hybrids.
It is also suitable for studying the mainstream layered materials, such as group-IV~\cite{NewJ.Phys.16(2014)125002J.Y.Wu, RSCAdv.5(2015)51912J.Y.Wu, Phys.Rev.B94(2016)045410S.C.Chen, Phys.Rev.B94(2016)205427J.Y.Wu}, group-V~\cite{J.Phys.Soc.Jpn.50(1981)3362Y.Takao, PhysicaBandC105(1981)93Y.Takao, Phys.Rev.B91(2015)085409E.T.Sisakht, Phys.Rev.B95(2017)115411J.Y.Wu}, and TMD~\cite{Phys.Rev.88(2013)075409E.Cappelluti, AIPAdv.3(2013)052111F.Zahid, Phys.Rev.B89(2014)155316Y.H.Ho, Appl.Phys.Lett.105(2014)222411Y.H.Ho, J.Phys.-Condens.Matter27(2015)365501E.Ridolfi, RSCAdv.5(2015)20858Y.H.Ho} 2D materials.

%-------------------------------------
%From [1607.02712v1]
%This model has been widely adopted to make systematic studies on multi-dimensional carbon-based materials, ranging from three-dimensional (3D) graphites [REF 1-6], 2D graphenes [REF 7-12], and 1D graphene nanoribbons [REF 13-15].
%It is also suitable for studying the mainstream layered materials, such as group-IV and group-V [REF J.Y. Wu, bilayer P] 2D materials [REF 16-19], and MoS2 [REF 20-22].
%-------------------------------------

%-------------------------------------
%CNT citations in GNR REVIEW
%323 326-329 332 334
%322-324 325
%82 84 85 299
%262  263
%-------------------------------------

%-----------------------------------
%
%4th SUBSECTION
%
%-----------------------------------

In this work, buckled monolayer GaAs with each atom being passivated by a F atom is chosen as a model study [Fig.~\ref{fig:2D_Geometry_BS_WF}(a)].
%Both $sp^3$-hybridized Ga and As atoms bond to four atoms (three are As or Ga, one is F) and thus free from dangling bonds and eliminate the effect of substrate.
The dangling bonds are saturated, and thus the effects of substrate can be eliminated.
The generalized tight-binding model, simultaneously considering geometric structure, mutli-orbital hybridization, SOC, and external fields, is employed to explore the magneto-electronic properties.
The low-lying electronic structure is composed of a direct energy gap and three groups of SOC-induced spin-polarized subbands presenting monotonous energy dispersions with strong wavevector-dependent ($\mathbf{k}$-dependent) spin splitting.
The state probabilities giving the detailed informations about the dominated/minor orbitals of each subband and their $\mathbf{k}$-dependent variations are discussed.
Magnetic quantization, accumulating electronic states with similar energies, induces three groups of LLs.
The initial energy of each LL group, subband degeneracy, energy spacing among LLs, and spin polarization are investigated.
The LL state probabilities, whose oscillation patterns are similar to those of harmonic oscillators with regular nodes at the localization centers, are analyzed.
The complex variation of LL domiated/minor orbitals are observed to reflect the average of accumulated neighboring zero-field electronic states.
It is predicted that the LL energies have the linear-$B_z$ dependence with the enhancement of spin splitting for an increasing magnetic field.
The given density of states (DOS) directly reflects the main features of the LL energy spectra in the structure, height, number, and frequency of the three-group spin-polarized LL peaks.
The electric field, contributing to an electric potential difference in the buckled structure, gives rise to monotonous/nonmonotonous energy dispersions, LL crossing, gap modulation, and enhancement of spin splitting.
The complex gap modulations and phase transitions based on the competition between magnetic and electric fields are investigated in detail.
A phase diagram about the complex phase transitions between four characteristic regions is illustrated, presenting that the external-fields-controlled gap presents several types of modulation, associated to different region-to-region phase transitions.
A brief comparison between the buckled monolayer GaAs and planer graphene is described for the differences in essential properties and responses to external fields based on the orbital domination, SOC, and geometric structure.
The predicted magneto-electronic properties of the monolayer GaAs, including three groups of spin-polarized LL DOS peaks with linear $B_z$-dependence, SOC-induced spin splitting, the external-field-controlled gap modulation/phase transition, electric-field-enhanced spin splitting, could be identified by scanning tunneling spectroscopy (STS) measurements.
Furthermore, this work can serve as a model study for understanding magnetic quantizations of other group III-V 2D materials.

\section{Generalized tight-binding model}

%!!!!!!!!!!!!!!!!!!!!!!!!!!!
%pz states to deep energy, far from EF

%Geometric structure
Monolayer GaAs has the buckled honeycomb lattice with each atom being passivated by a F atoms [Fig.~\ref{fig:2D_Geometry_BS_WF}(a)].
%The space group of the film is p3m1 (no.156).
Both $sp^3$-hybridized Ga and As atoms bond to four atoms (three for As or Ga; one for F) and the Ga-As bond length is about 2.521 {\AA}.
%, analogous to the case of the bulk counterparts.
%However, the $T_d$ symmetry along the [111] direction of GaAs crystal is broken in this GaAs film.
%There is only $C_{3v}$ symmetry in this film.
%The Ga-As bond length is about 2.521 {\AA}, slightly longer than that of GaAs crystal 2.489 {\AA}.
%The optimized lattice constant (the length of base vectors) is 4.226 {\AA}.
%The primitive unit vectors, $\mathbf{a}_1$ and $\mathbf{a}_2$ along the two sides of rhombus, with a lattice constant of $a = 4.226$ {\AA}.
A unit cell containing two different Ga and As sublattices is indicated by the rhombus with the primitive unit vectors, $\mathbf{a}_1$ and $\mathbf{a}_2$ of a lattice constant $a = 4.226$ {\AA}.
%Each unit cell is indicated by the rhombus with the primitive unit vectors, $\mathbf{a}_1$ and $\mathbf{a}_2$ of a lattice constant $a = 4.226$ {\AA}.
%Each unit cell contains two different Ga and As sublattices, and
The altitude of the buckled structure measured from the distance between the Ga-plane and As-plane is $l_z = 0.633$ {\AA} [Fig.~\ref{fig:2D_Geometry_BS_WF}(b)].
The buckling angle $\theta$ between the Ga-As bond and the $z$-axis is about $104.54^\circ$.
%F atoms are right above (below) the Ga (As) atoms with the Ga-F and As-F distances of 1.776 {\AA} and 1.781 {\AA}, respectively.
%The binding energies of the Ga-F and As-F bonds are 25.73 eV and 25.38 eV, suggesting that F atoms are chemically bonded to the monolayer GaAs.
This configuration is free from dangling bonds and thus chemically stable~\cite{Sci.Rep.5(2015)8441M.W.Zhao}.
%
%
%
%
%
%From the experimental point of view, the growth of GaAs ultrathin films has been achieved on the substrates, such as silicon (111) surface~\cite{Phy.-Usp.51(2008)437B.B.Yu}, and fluorination of GaAs can be achieved at low temperature in $\mathrm{CF}_4$ plasma~\cite{Jpn.J.Appl.Phys.30(1991)1581M.Iida}. Benefiting from the well-developed GaAs technology and recent progresses in nanotechnology, the realization of the fluorinated GaAs film seems plausible in the near future.
%

%Tight-binding model
To illustrate the electronic properties explicitly, the Hamiltonian built from the  tight-binding functions of $4s$, $4p_x$, and $4p_y$ orbitals is expressed as
\begin{equation}
\mathcal{H} = \sum_{m,\alpha} \epsilon_m^\alpha c_m^{\alpha \dagger} c_m^{\alpha} +
       \sum_{\langle m,n \rangle ,\alpha ,\beta} \gamma_{mn}^{\alpha \beta}
       \big( c_m^{\alpha \dagger} c_m^{\beta} + h.c. \big),
\label{eq:Hamiltonian_monoGaAs}
\end{equation}
where $\epsilon_m^\alpha$, $c_m^{\alpha \dagger}$, and $c_m^{\alpha}$ respectively represent the on-site energy, creation, and annihilation operators of an electron at the $\alpha$-orbital of the $m$-th atom.
$\gamma_{mn}^{\alpha \beta}$ is the nearest-neighbor hopping integral between an $\alpha$-orbital of the $m$-th atom and a $\beta$-orbital of the $n$-th atom.
The multi-orbital hopping integrals are
%$\gamma_{ij}^{ss} = V_{ss\sigma}$,
%$\gamma_{ij}^{sp_x} = V_{sp\sigma} \times \cos\theta$,
%$\gamma_{ij}^{sp_y} = V_{sp\sigma} \times \cos\phi$,
%$\gamma_{ij}^{p_xp_x} = V_{pp\sigma} \times \cos^2\theta + V_{pp\pi} \times (1-\cos^2\theta)$,
%$\gamma_{ij}^{p_yp_y} = V_{pp\sigma} \times \cos^2\phi + V_{pp\pi} \times (1-\cos^2\phi)$,
%$\gamma_{ij}^{p_xp_y} = (V_{pp\sigma} - V_{pp\pi}) \times \cos\theta \times \cos\phi$,
%$\gamma_{mn}^{ss} = V_{ss\sigma}$,
%$\gamma_{mn}^{sp_x} = V_{sp\sigma} \cos\theta$,
%$\gamma_{mn}^{sp_y} = V_{sp\sigma} \cos\phi$,
%$\gamma_{mn}^{p_xp_x} = V_{pp\sigma} \cos^2\theta + V_{pp\pi} (1-\cos^2\theta)$,
%$\gamma_{mn}^{p_yp_y} = V_{pp\sigma} \cos^2\phi + V_{pp\pi} (1-\cos^2\phi)$,
%and $\gamma_{mn}^{p_xp_y} = (V_{pp\sigma} - V_{pp\pi}) \cos\theta \cos\phi$,
%where $\theta$ and $\phi$ are the angles of the vector pointed from the $m$-th atom to the $n$-th atom with respect to the $x$- and $y$-axis~\cite{Phys.Rev.94(1954)1498J.C.Slater}, and the Slater-Koster hopping parameters in the $sp^3$ bonding optimized at the equilibrium state are $V_{ss\sigma} = -1.707$ eV, $V_{sp\sigma} = 2.056$ eV, $V_{pp\sigma} = 2.650$ eV, and $V_{pp\pi} = -0.827$ eV~\cite{Sci.Rep.5(2015)8441M.W.Zhao}.
$\gamma_{mn}^{ss} = V_{ss\sigma}$,
$\gamma_{mn}^{sp_x} = V_{sp\sigma} \cos\theta_x$,
$\gamma_{mn}^{sp_y} = V_{sp\sigma} \cos\theta_y$,
$\gamma_{mn}^{p_xp_x} = V_{pp\sigma} \cos^2\theta_x + V_{pp\pi} (1-\cos^2\theta_x)$,
$\gamma_{mn}^{p_yp_y} = V_{pp\sigma} \cos^2\theta_y + V_{pp\pi} (1-\cos^2\theta_y)$,
and $\gamma_{mn}^{p_xp_y} = (V_{pp\sigma} - V_{pp\pi}) \cos\theta_x \cos\theta_y$,
where $\theta_x$ and $\theta_y$ are respectively the angles of the vector pointed from the $m$-th atom to the $n$-th atom with respect to the $x$- and $y$-axis~\cite{Phys.Rev.94(1954)1498J.C.Slater}, and the Slater-Koster hopping parameters in the $sp^3$ bonding optimized at the equilibrium state are $V_{ss\sigma} = -1.707$ eV, $V_{sp\sigma} = 2.056$ eV, $V_{pp\sigma} = 2.650$ eV, and $V_{pp\pi} = -0.827$ eV~\cite{Sci.Rep.5(2015)8441M.W.Zhao}.
The on-site energies of $s$- and $p$-orbitals are set to the values ($-12.00$ eV, $-5.67$ eV) for Ga and ($-17.68$ eV, $-8.30$ eV) for As, being taken from those of bulk GaAs~\cite{J.Phys.Chem.Solids44(1983)365P.Vogl}.
%Other parameters optimized at the equilibrium state are $V_{ss\sigma} = -1.707$ eV, $V_{sp\sigma} = 2.056$ eV, $V_{pp\sigma} = 2.650$ eV, and $V_{pp\pi} = -0.827$ eV, respectively~\ref{Sci.Rep.5(2015)8441M.W.Zhao}.

%%Bridge paragraph
%It is noteworthy that both GaF and SnF films have structure inversion symmetry, whereas the fluorinated GaAs hasn’t. In this sense, fluorinated GaAs represents a more general model system of buckled honeycomb lattices. Some exotic physical phenomena arising from the breakage of inversion symmetry, such as Rashba and Dresselhaus spin-orbital coupling, appear in the GaAs film as described below.

%Spin-orbit interaction
%Such spinsplitting has also been found in GaAs quantum wells25, where the lifting of spin degeneracy due to SOC leads to terms linear in electron wave vector k in the effective Hamiltonian26. The origin of the linear terms in low-dimensional systems is structure inversion asymmetry which lead to Rashba and Dresselhaus spin-orbital terms in the Hamiltonian26,27.

When an electron with momentum $\mathbf{p}$ moving close to the atomic nuclei in a crystal with potential $V$, it experiences an effective magnetic field $B_{\mathrm{eff}}\sim\nabla V\times\mathbf{p}/m_0c^{2}$ in its rest-frame ($m_0$ is the mass of a free electron and $c$ is the speed of light).
Such field induces a momentum-dependent Zeeman energy called the SO coupling, which is given by
\begin{equation}
H^{SO}=\frac{\hbar}{4m_0^{2}c^{2}}(\nabla V\times\mathbf{p})\cdot\mathbf{\boldsymbol{\sigma}},
\label{eq:Hso_monoGaAs}
\end{equation}
where $\hbar$ is the reduced Planck constant and $\boldsymbol\sigma$ is the vector of Pauli matrices.
In the central field approximation, the crystal potential $V(\mathbf{r})$ is considered as the spherical atomic potential.
The SOC term on the same atom is taken into account and it can be obtained by calculating the mean value:
\begin{equation}
H_{i,\alpha\beta}^{SO}=\lambda_{i}\langle\mathbf{L}\cdot\boldsymbol{\sigma}\rangle_{\alpha\beta},
\label{eq:Hso_element_monoGaAs}
\end{equation}
where $\lambda_{i}$ is the SOC strength of the $i$-th atom and $\mathbf{L}$ is the orbital angular momentum operator.
The matrix element $\langle\mathbf{L}\cdot\boldsymbol{\sigma}\rangle_{\alpha\beta}$ is given in the basis of atomic orbitals $(\alpha,\beta)$, and the dimensionless SOC operator $\mathbf{L}\cdot\boldsymbol{\sigma}$ for the relevant orbitals ($4s$, $4p_x$, and $4p_y$) in the 2D system is given by
\begin{equation}
\mathbf{L} \cdot \boldsymbol{\sigma} = \left(\begin{array}{ccc}
0 & 0 & 0\\
0 & 0 & -\mathit{i}s_{z}\\
0 & \mathit{i}s_{z} & 0
\end{array} \right),
\label{eq:L_sigma_monoGaAs}
\end{equation}
where $s_{z}=\left(\begin{array}{cc}
1 & 0\\
0 & -1
\end{array}\right)$.
The SOC strengths of Ga and As atoms are chosen to be 0.058 eV and 0.140 eV, respectively~\cite{Rev.Mod.Phys.76(2004)323I.Zutic}.

%
%Our TB Hamiltonian is also available for understanding the band gap opening process by involving a spin-orbital component ($\mathcal{H}^{SOI}$).
%\begin{equation}
%\mathcal{H}^{SOI} = \bigg( \frac{\hbar}{4m^2c^2} \bigg) ( \nabla V \times \mathbf{p}) \cdot \sigma,
%\label{eq:H_SOI_monoGaAs}
%\end{equation}
%The major SOC comes from the orbits close to the atomic nuclei. Therefore, the crystal potential $V(\mathbf{r})$ can be approximated by the spherical atomic potential. By averaging the radial degree of freedom, it reads
%\begin{equation}
%\mathcal{H}^{SOI}_{i,\alpha \beta} = \lambda_i \langle \mathbf{L} \cdot \mathbf{\sigma} \rangle_{\alpha \beta},
%\label{eq:H_SOI2_monoGaAs}
%\end{equation}
%$\sigma$ is the vector of the of the Pauli matrices and $\mathbf{L}$ is the angular
%momentum operator. The matrix element, $\langle \mathbf{L} \cdot \mathbf{\sigma} \rangle_{\alpha \beta}$ is given in the basis of directed atomic orbitals (a, b) and li is the SOC strength of the $i$-th atom. The matrix elements of the dimensionless SOC operator $\mathbf{L} \cdot \mathbf{\sigma}$ for the relevant orbitals ($s$, $p_x$, and $p_y$) in the 2D system are:
%\begin{equation}
%\mathbf{L} \cdot \mathbf{\sigma} = \left(\begin{array}{ccc}
%0 & 0 & 0\\
%0 & 0 & -\mathit{i}s_{z}\\
%0 & \mathit{i}s_{z} & 0
%\end{array} \right),
%\label{eq:L_sigma_monoGaAs}
%\end{equation}
%The SOC strengths of Ga and As atoms are set to 0.058 eV and 0.140 eV, respectively~\ref{Rev.Mod.Phys.76(2004)323I.Zutic}.
%

%Peierls substitution
When a uniform perpendicular magnetic field, $\mathbf{B}=B_z \hat{z}$, is applied to monolayer GaAs, the effective Hamiltonian can be regarded as the Peierls substitution Hamiltonian~\cite{Z.Phys.80(1933)763R.Peierls}.
Each Hamiltonian matrix element turns into the product of the zero-field element and the extra Peierls phase, $\exp (i2\pi \theta _{mn})$, where $\theta_{mn}=(1 / \phi_{0}) \int_{m}^{n}\mathbf{A}\cdot d\mathbf{l}$ is a line integral of the vector potential $\mathbf{A}$ from the $m$-th to $n$-th site, $\mathbf{A}$ is chosen as $(0,B_z x,0)$ in the Landau gauge, and $\phi_0=h/e$ ($4.1357 \times 10^{-15}$ T$\cdot$m$^2$) is the magnetic flux quantum~\cite{Superfluids1(1950)152F.London, Proceedings(1953)935L.Onsager}.
%F. London, Superfluids (John Wiley & Sons, New York, 1950), p. 152.
%L. Onsager, Proceedings of the International Conference on Theoretical Physics, Kyoto and Tokyo, September, 1953 (Science Council of Japan, Tokyo, 1954), pp. 935-6.
%----------------------------------
%Magnetic flux quantum of superconductoring rings
%Phys. Rev. Lett. 7 (1961) 43, B.S. Deaver
%Phys. Rev. Lett. 7 (1961) 51, H. Doll
%----------------------------------
%Here, $\mathbf{A}$ is chosen as $(0,B_z x,0)$ to preserve the translational invariance in the $x$-direction under the Landau gauge, and $\phi_0=h/e$ is the magnetic flux quantum.
%The unit cell becomes an enlarged rectangle with $2R_B$ Ga and $2R_B$ As atoms to satisfy the periodicity of Peierls phase, where $R_B = 26739/B_z$ is the ratio of $\phi_0$ to magnetic flux through a hexagon [Fig.~\ref{fig:Geo_LL_WF_monoGaAs}(a)].
The unit cell becomes an enlarged rectangle with $2R_B$ Ga and $2R_B$ As atoms to satisfy the periodicity of Peierls phase, where $R_B = \phi_0 / \phi = \phi_0 / (B_z \sqrt{3}/2a^2) \sim 26739~\mathrm{T}/B_z$ is the ratio of flux quantum to magnetic flux through a hexagon $\phi$ [Fig.~\ref{fig:Geo_LL_WF_monoGaAs}(a)].
The reduced Brillouin zone has an area of $4\pi^2/\sqrt{3}a^2R_B$.
%The $B_z$-dependent reduced Brillouin zone has an area of $4\pi^2/\sqrt{3}a^2R_B$.
The Hamiltonian is built in the space spanned by the $24R_B$ tight-binding functions
%$\{ s^{Ga}, p_x^{Ga}, p_y^{Ga}, s^{As}, p_x^{As}, p_y^{As} \} \bigotimes \{\uparrow , \downarrow \}$
%$\{ | Ga_m^o \rangle , | As_m^o \rangle ; m=1,2,3,...,2R_B; o=4s,4p_x,4p_y \} \bigotimes \{\uparrow , \downarrow \}$.
$\{ | Ga_m^{orb} \rangle , | As_m^{orb} \rangle ; m=1,2,3,...,2R_B; orb=4s,4p_x,4p_y \} \bigotimes \{\uparrow , \downarrow \}$.
An electric field $\mathbf{E} = E_z \hat{z}$ along the $z$-axis introduces a potential energy $-e E_z l_z/2$ ($e E_z l_z/2$) to the site energy of the Ga (As) sublattice.
The exact diagonalization of the Hamiltonian matrix $\mathcal{H}$ yields the energy spectrum $E^{c,v}$ and wave functions $| \Psi^{c,v} \rangle$, where the superscripts $c$ and $v$ denote the conduction and valence subbands, respectively.
The generalized tight-binding model can be further developed to comprehend the Landau quantization in other layered systems with complex orbital bondings and spin configurations.
%The generalized tight-binding model can be further developed to comprehend the magnetic quantization in other layered systems with complex orbital bonding and spin configurations.

\section{Spin-polarized magneto-electronic properties}

\begin{figure*}
  \includegraphics[]{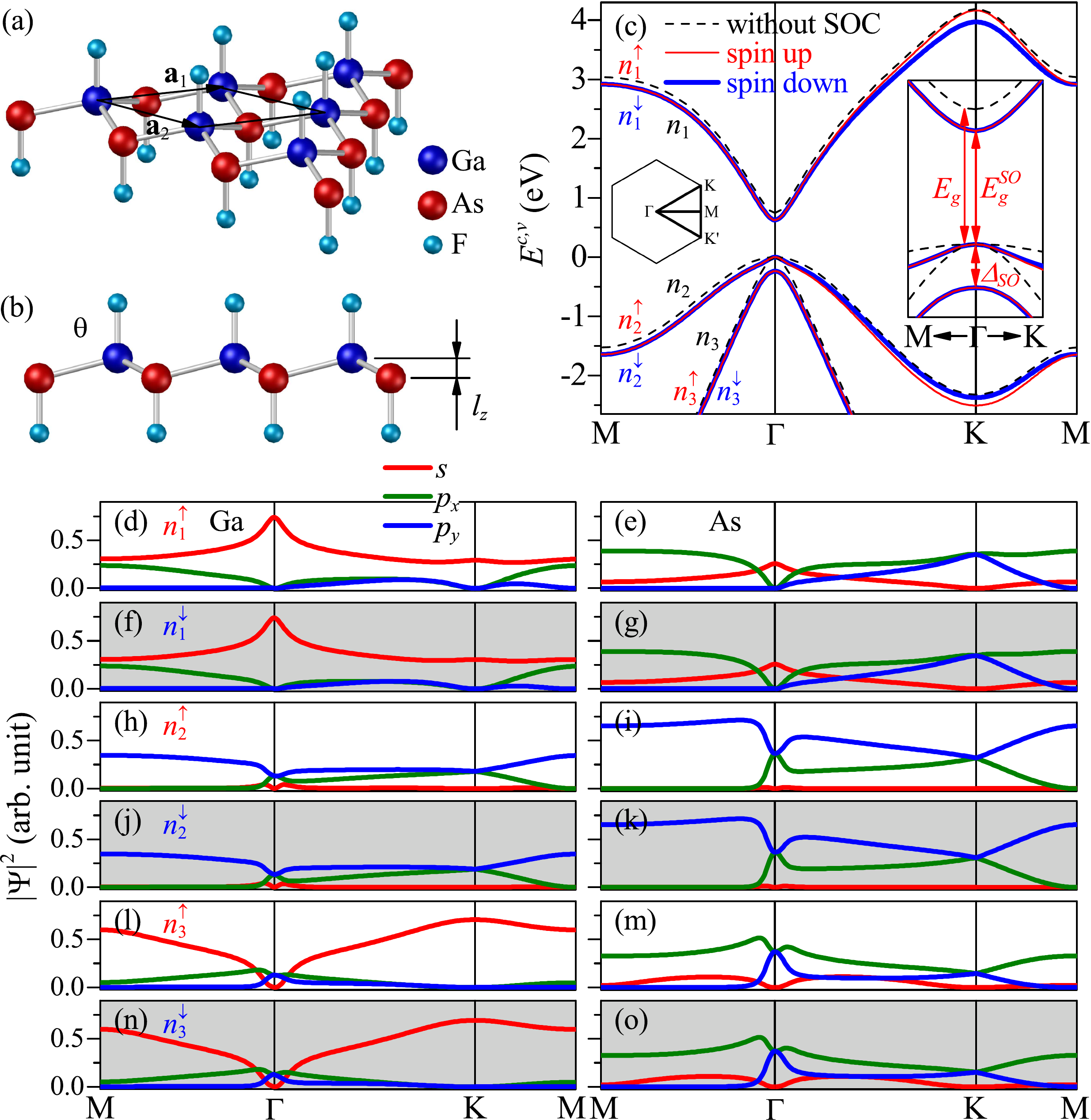}\\
  \caption{
(Color online)  
(a) Schematic representation of the monolayer GaAs decorated by F atoms.
The unit cell is indicated by the rhombus.
$\mathbf{a}_1$ and $\mathbf{a}_2$ are the two translation vectors.
(b) Side view of the low-buckled monolayer GaAs with the buckling angle $\theta$.
(c) Spin-degenerate energy subbands without SOC ($n_1$, $n_2$, and $n_3$) and SOC-induced spin-polarized subbands ($n_1^\uparrow$, $n_1^\downarrow$, $n_2^\uparrow$, $n_2^\downarrow$, $n_3^\uparrow$, and $n_3^\downarrow$) along the high symmetry points.
(d)--(o) State probabilities of various orbitals located at two sublattices Ga and As with spin-up (white zones) and spin-down (gray zones) arrangements.
$s$, $p_x$, and $p_y$ orbitals are indicated by red, green, and blue curves, respectively.
}
  \label{fig:2D_Geometry_BS_WF}
\end{figure*}

Monolayer GaAs has feature-rich energy bands, mainly owing to the low-buckled structure, $sp^3$ bonding, and SOC.
There exist three low-lying energy subbands, i.e., the unoccupied conduction subband ($n_1$) and two occupied valence subbands ($n_2$ and $n_3$) with different curvatures near the $\Gamma$ point, touching at the Fermi level ($E_F = 0$).
Without the SOC, they have the strong wavevector-dependence in the monotonous form [dashed curves in Fig.~\ref{fig:2D_Geometry_BS_WF}(c)].
Each subband is two-fold degenerate for the spin degree of freedom except that the $n_2$ and $n_3$ subbands intersect and possess a four-fold degeneracy.
The conduction and valence subbands near the $\Gamma$ point are respectively dominated by the $4s$ and $(4p_x, 4p_y)$ orbitals~\cite{Sci.Rep.5(2015)8441M.W.Zhao}.
More importantly, a direct band gap of $E_g = 0.742$ eV is determined by the band-edge states of $n_1$ and $n_2$/$n_3$ at the $\Gamma$ point.
%SOC further induces the variation of band gap, split of subbands, lift of state degeneracy, and spin splitting [solid curves in Fig.~\ref{fig:2D_Geometry_BS_WF}(c)].
The significant SOC further induces the variation of band gap and spin splitting [solid curves in Fig.~\ref{fig:2D_Geometry_BS_WF}(c)].
%SOC further induces the variation of band gap and spin splitting [solid curves in Fig.~\ref{fig:2D_Geometry_BS_WF}(c)].
%The band gap shrinks to $E_g^{SO} = 0.623$ eV, while the $n_2$ subband is separated from the $n_3$ subband by $\Delta_{SO} = 0.237$ eV, lifting the state degeneracy at the $\Gamma$ point from four- to two-fold.
The band gap shrinks to $E_g^{SO} = 0.623$ eV, while the $n_2$ and $n_3$ valence subbands are separated by $\Delta_{SO} = 0.237$ eV, lifting the state degeneracy at the $\Gamma$ point from four- to two-fold.
%spin-orbit splitting of the valence band $\Delta_{SO} = 0.237$ eV
%The spin degeneracy is removed at the zone except the $\Gamma$ point.
The spin degeneracy is removed except for the zone from the $\Gamma$ point to the M point.
Therefore, the spin-degenerate subbands become spin-polarized subbands.
The splitting energies between spin-up and spin-down subbands gradually increase when deviating from the $\Gamma$ point and reach maxima at the K (K') points (e.g., $0.196$ eV between $n_1^\uparrow$ and $n_1^\downarrow$ subbands; $0.133$ eV between $n_2^\uparrow$ and $n_2^\downarrow$ subbands).
%[OPPOSITE spin up and down states at K and K']
Such spin splitting has also been found in GaAs quantum wells by photocurrent measurements~\cite{Phys.Rev.B75(2007)035327S.Giglberger}, where SOC leads to terms linear in wavevector $\mathbf{k}$ in the effective Hamiltonian~\cite{J.Phys.C17(1984)6039Y.A.Bychkov}.
%The origin of the linear terms in low-dimensional systems is structure inversion asymmetry which lead to Rashba and Dresselhaus spin-orbital terms in the Hamiltonian26,27.
%[NEED to realize]The spin splitting is crucial for the field of spintronics, indeed it allows the electric field control of spin polarization, determines the spin relaxation rate, and can be utilized for all-electric spin injection~\cite{Rev.Mod.Phys.76(2004)323I.Zutic}.

%From [Phys. Rev. B 75 (2006) 035327, S. Giglberger]
%In low dimensional structures based on III-V compound semiconductors the spin degeneracy of the energy bands is removed. This lifting of spin degeneracy is caused by spinorbit interaction and results in terms linear in electron wave vector k in the effective Hamiltonian. The spin splitting is crucial for the field of spintronics, indeed it allows the electric field control of spin polarization, determines the spin relaxation rate, and can be utilized for all-electric spin injection~\cite{Rev.Mod.Phys.76(2004)323I.Zutic}.

%ADD  $\mathbf{a}_1$ and $\mathbf{a}_2$

%-------------------------------------

%[State probability at zero field]

%-------------------------------------

The state probability ($| \Psi^{c,v} |^2$) exhibits the spacial contribution of different orbitals on subbands and figures out the variations of dominated/minor orbitals near various high-symmetric points.
A whole range of the orbital variation on different sublattices for the  $n_1^{\uparrow \downarrow}$, $n_2^{\uparrow \downarrow}$, and $n_3^{\uparrow \downarrow}$ subbands is shown in Figs.~\ref{fig:2D_Geometry_BS_WF}(d)--(o).
%Despite the fact of spin splitting in energy bands, the state configurations of spin-up and spin-down states are quite similar.
%It is sufficient to discuss one of the polarized states (e.g., white zones).
It is sufficient to discuss one of the polarized states (e.g., spin-up states) because the state configurations of spin-up (white zones) and spin-down (gray zones) states are quite similar. % (despite the fact of spin splitting in energy bands)
%, where red, green, and blue curves indicate the $s$, $p_x$, and $p_y$ orbitals, respectively.
%It is sufficient to discuss one of the polarized states (e.g. white zones), where red, green, and blue curves indicate the $s$, $p_x$, and $p_y$ orbitals, respectively.
%[From arXiv] The first pair of valence and conduction bands have the doubly degenerate states associated with the spin-down- and spin-up-dominated equivalent configurations. It is sufficient to only discuss one of both configurations, as shown in Figs. 7(b) and 7(c) for germanene.
The state probabilities for different orbitals are very sensitive to the sublattices and wavevectors.
%[For example,...  Ga =/ As  ]
%Gamma -- K px py deviate
%Gamma and K, px py merge
%In addition, G K M = G K' M
In the conduction $n_1^{\uparrow \downarrow}$ and valence $n_3^{\uparrow \downarrow}$ subbands, the $s$-orbitals (red curves) and $p_x$-orbitals (green curves) are respectively the most dominated contributions on the Ga and As sublattices for a wide range of $\mathbf{k}$.
The $p_y$-orbitals (blue curves) are the dominated contributions in the valence $n_2^{\uparrow \downarrow}$ subbands.
Remarkably, the $p_x$- and $p_y$-orbitals on a specific sublattice are of identical intensity at the high-symmetric $\Gamma$ and K points.
%The state probabilities near the $\Gamma$ point, revealing the orbital variation for the low-lying states, are much different from the probabilities far away the $\Gamma$ point.
The state probabilities near the $\Gamma$ point, which are much different from the probabilities far away the $\Gamma$ point, reveal the orbital variation for the low-lying states.
The conduction subbands are dominated by the $s$-orbitals, whose state probabilities on the Ga sublattice is larger than those on the As sublattice [Figs.~\ref{fig:2D_Geometry_BS_WF}(d)--(g)].
The increase of $p_x$- and $p_y$-orbital strength and the decrease of $s$-orbital strength arise as $\mathbf{k}$ deviates from the $\Gamma$ point.
%The increase of $p_x$- and $p_y$-orbital strength and the decrease of $s$-orbital strength arise as $\mathbf{k}$ deviates from $(k_x, k_y) = (0, 0)$.
The valence $n_2$ ($n_3$) subbands are dominated by $p_y$-orbitals ($p_x$-orbitals) [Figs.~\ref{fig:2D_Geometry_BS_WF}(h)--(o)].
Instead of the Ga sublattices, the dominated orbitals on the As sublattices possess larger strength.
It should be noted that the relative strength of the orbital probabilities for the low-lying states will reflect on the quantized magneto-electronic states.
In other words, the low-energy Landau states features those accumulated zero-field states near the $\Gamma$ point (discussed later).

\begin{figure*}
  \includegraphics[]{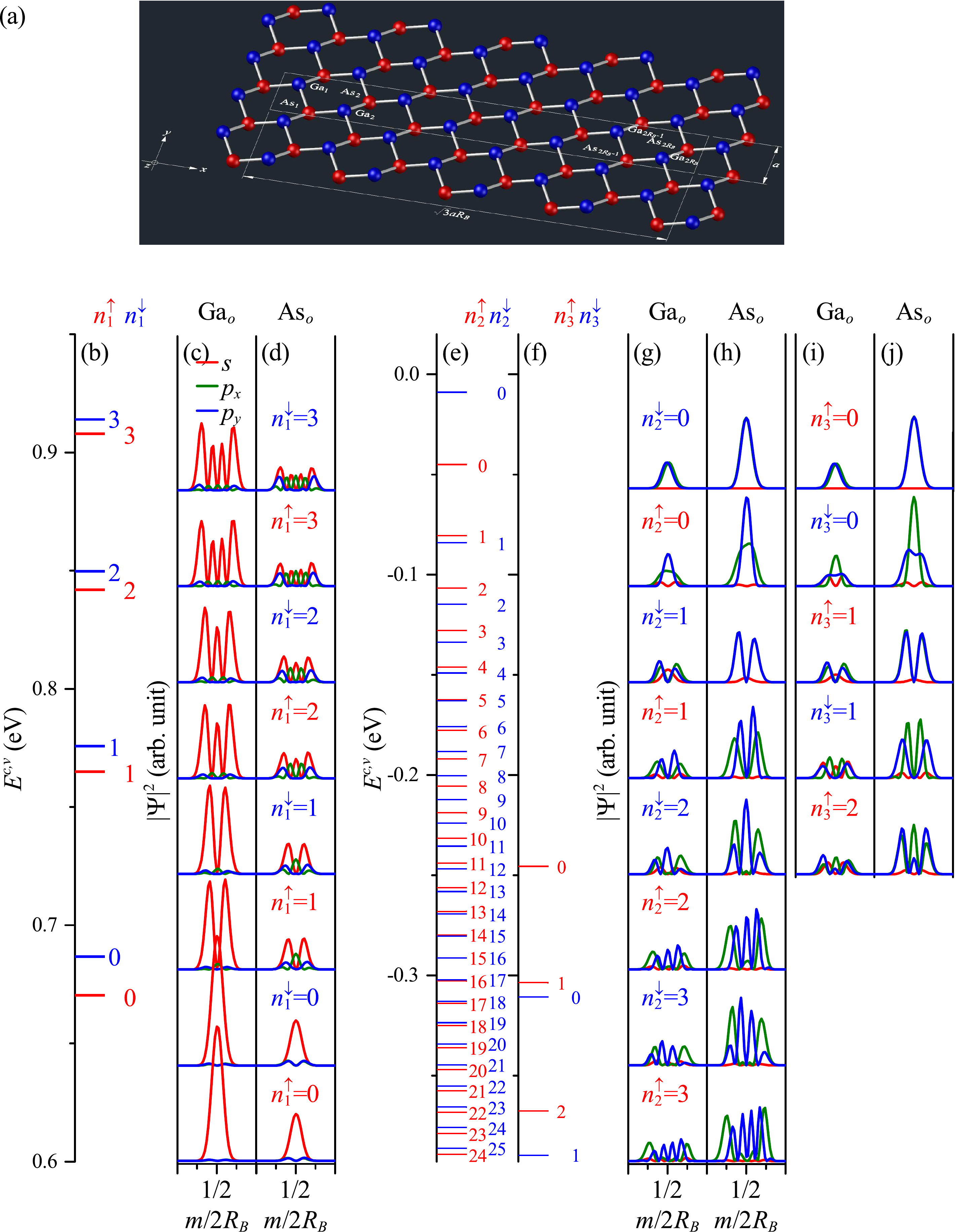}\\
  \caption{
(Color online)    
(a) Geometric structure of the low-buckled monolayer GaAs with an enlarged rectangular unit cell in $B_z\hat{z}$.
%(b)--(j) LL energies of various groups and the corresponding probabilities of the subenvelope functions near the localization center at $B_z = 60$ T.
(b)--(j) Spin-polarized LLs and the corresponding probabilities of the subenvelope functions near the localization center at $B_z = 60$ T.
}
  \label{fig:Geo_LL_WF_monoGaAs}
\end{figure*}

Magnetic fields constrain carrier motions in real space, bring neighboring electronic states together, and induce highly degenerate Landau states.
Near the Fermi energy, there are three groups of spin-polarized dispersionless LLs, i.e., one group of occupied conduction LLs [$n_1^{\uparrow}$ and $n_1^{\downarrow}$ in Fig.~\ref{fig:Geo_LL_WF_monoGaAs}(b)] and two groups of unoccupied valence LLs [$n_2^{\uparrow}$, $n_2^{\downarrow}$, $n_3^{\uparrow}$, and $n_3^{\downarrow}$ in Figs.~\ref{fig:Geo_LL_WF_monoGaAs}(e) and (f)].
%Magnetic fields quantize the electronic states of similar energies and induce three groups of low-lying spin-polarized dispersionless Landau levels (LLs), i.e., one group of occupied conduction LLs [$n_1^{\uparrow}$ and $n_1^{\downarrow}$ in Fig.~\ref{fig:Geo_LL_WF_monoGaAs}(b)] and two groups of unoccupied valence LLs [$n_2^{\uparrow}$, $n_2^{\downarrow}$, $n_3^{\uparrow}$, and $n_3^{\downarrow}$ in Figs.~\ref{fig:Geo_LL_WF_monoGaAs}(e) and (f)].
%Owing to the SOC between the $4p_x$ and $4p_y$ orbitals, three groups of LLs are distinctly spin-polarized.
The distinct spin polarization in each group of LLs results from the SOC between the $4p_x$ and $4p_y$ orbitals.
%Vsoc between the $4p_x$ and $4p_y$ orbitals leads to distinct spin-polarized LLs
Their LL initial energies are respectively near $0.62$ eV, $0$ eV, and $-0.24$ eV, which reflect the energies of electronic states at the $\Gamma$ point in the absent of magnetic fields.
For each ($k_x, k_y$), all LLs are two-fold degenerate, being attributed to the one $\Gamma$-valley degree of freedom and the mirror symmetry of $z = 0$ plane.
As the state energy grows, the energy spacing between LLs of the same spin-up (or spin-down) subgroup gradually shrinks.

%The conduction LLs initiate from $\sim 0.62$ eV (Fig.~\ref{fig:Geo_LL_WF_monoGaAs}).
%!!!!!!!!!!!!!!!!The spin-down LLs possess higher energies than the spin-up LLs.
%The valence heavy and light hole LLs initiate from $\sim 0$ eV and $\sim -0.24$ eV, respectively.
%The spin-up and spin-down HH and LH LLs seems don't have an apparent regularity.
%The initial energies of various LL groups close to the energies of zero-field subband at the $\Gamma$ point.
%The spin-polarized feature and the initial energies of various LL groups, which reflect the main features of zero-field band structure near the $\Gamma$ point.

%SP of LL

%State probabilities, which present the spatial information of electronic states, are very significant for realizing fundamental physical properties, such as charge densities,[!124-126] state mixing,[!127] and optical selection rules.[!65,78]
%, which is useful for understanding other essential properties, such as optical absorption spectra and selection rules.
Wave functions, presenting the spatial information of electronic states, are very important in realizing fundamental physical properties, such as charge densities~\cite{Phys.Rev.28(1926)1049E.Schrodinger, Phys.Rev.B78(2008)235311M.W.Y.Tu, Phys.Rev.B89(2014)121401P.M.PerezPiskunow, Phys.Rev.B90(2014)115423G.Usaj}, state mixing~\cite{Philos.Mag.94(2014)1859H.C.Chung, Phys.Chem.Chem.Phys.15(2013)868H.C.Chung}, and optical selection rules~\cite{Opt.Express19(2011)23350H.C.Chung, Phys.Chem.Chem.Phys.18(2016)7573H.C.Chung, Carbon109(2016)883H.C.Chung, Phys.Rev.B95(2017)155438V.A.Saroka}.
Under the influence of magnetic fields, wave functions in monolayer GaAs present peculiar spatial distributions, where the localization center, orbital domination, waveform, and node number are very sensitive to the wavevector, state energy, and spin polarization.
%Each spin-polarized LL wave function is characterized by the subenvelope functions on the different sublattices with the $sp^3$ orbitals.
%WF decomposition [formula] Especially, in ZGNRs, the wave functions can be decomposed into the subenvelope functions on the Ga and As sublattices at the odd and even sites.
%For simplicity , WF2 are discussed
%the probability of the subenvelope functions
%the evolution of distribution probabilities of subenvelope functions
Each spin-polarized LL wave function can be decomposed into subenvelope functions with the ($s$, $p_x$, $p_y$) orbitals on the Ga and As sublattices at the odd and even sites.
%For the sake of simplicity, only the distribution probabilities of subenvelope functions at the odd sites will be considered because those at the even sites have the same behavior.
For the sake of simplicity, only the distribution probabilities of subenvelope functions at the odd sites ($\mathrm{Ga}_o$ and $\mathrm{As}_o$) will be considered because the even-site probabilities have the same behavior as the odd-site ones.
The localization centers of the LL wave functions are strongly dependent on the wavevector.
%[]
%At ($k_x = 0, k_y = 0$),
At $(k_x, k_y) = (0, 0)$, one of the doubly degenerate spin-polarized LL states is localized at the $1/2$ position of the enlarged unit cell ($m/2R_B = 1/2$) and the other is localized at the 0 position.
%the degenerate states are localized at the $0$ and $1/2$ positions of the enlarged unit cell.
The Landau states at the $1/2$ position are chosen for illustrating the main features, since the state probabilities at $0$ and $1/2$ positions only differ in the localization center.
%The Landau state at the 0 position can be regarded as that at the 1/2 positions under a position shift of half enlarged unit cell. (length of half enlarged unit cell)
%The 1/2 states are chosen to illustrate the main features of the LL wave functions.
The probabilities of subenvelope functions are well-behaved in their spatial distributions.
Their oscillation patterns at the localization center are similar to those of harmonic oscillators, having regular node (zero-point) numbers.
%Their oscillation patterns at the localization center are similar to those of a harmonic oscillator, having regular node (zero-point) number for characterizing LLs.
%The state probabilities exhibit regular Landau modes for various orbitals, and the dominated orbital is appropriate for labeling LLs.
%The dominated orbital is significant for judging the characteristic of the Landau state and appropriate for labeling the state.
For any particular LL, the node numbers of various orbital subenvelope functions are identical for the Ga and As sublattices.
%The Ga and As sublattices present the same quantum mode after magnetic quantization.
The dominated-orbital subenvelope function is significant for characterizing the Landau state, and its node number, which gradually grows as the state energy increases, is appropriate for labeling the LL.
%As the state energy increases, the node number of dominated-orbital subenvelope function gradually grows, which is suitable for labeling the LLs.
%[[As a result of the hexagonal lattice, the nearest-neighbor atomic interactions near the $\Gamma$ point are roughly proportional to $k^2$. The Ga and As sublattices present the same quantum mode after magnetic quantization. !!! Is the $k^2$ BS at the $\Gamma$ point instead of the interactions]]
%In a Landau state, the node number for the probabilities of s-orbital (or px-, py- orbitals) subenvelope function in different sublattices are identical owning to the $k^2$-dependent energy dispersion at the $\Gamma$ point.
In the $n$th conduction LLs ($n_1^{\uparrow}=n$ and $n_1^{\downarrow}=n$), the probabilities of dominated $s$-orbital subenvelope functions with $n$ nodes have strength larger than the probabilities of minor ($p_x$ and $p_y$)-orbital subenvelope functions with $n+1$ nodes [Figs.~\ref{fig:Geo_LL_WF_monoGaAs}(c) and (d)].
%The $n$th conduction LLs ($n_1^{\uparrow}=n$ and $n_1^{\downarrow}=n$) are $s$-orbital-dominated,
In the $n_2^{\uparrow}=n$ ($n_2^{\downarrow}=n$) valence LLs, there are $n$, $n$, and $n+1$ ($n-1$) nodes in the dominated $p_y$-orbitals and minor $p_x$- and $s$-orbitals, respectively [Figs.~\ref{fig:Geo_LL_WF_monoGaAs}(g) and (h)].
In the $n_3^{\uparrow}=n$ ($n_3^{\downarrow}=n$) valence LLs, $n$, $n$, and $n-1$ ($n+1$) nodes are respectively in the dominated $p_x$-orbitals and minor $p_y$- and $s$-orbitals [Figs.~\ref{fig:Geo_LL_WF_monoGaAs}(i) and (j)].
It is noteworthy that the Landau state reflects the average of accumulated neighboring zero-field electronic states with similar energies.
In other words, the relative strength among LL subenvelope function probabilities of various orbitals (or different sublattices) corresponds to the relative strength among zero-field probabilities of various orbitals (or different sublattices).
In each LL group, the energy-dependent relative orbital strength is roughly associated with the $\mathbf{k}$-dependent relative orbital strength at $B_z = 0$ owing to the monotonously varying zero-field band structure near the $\Gamma$ point.
The dominated $s$-orbitals on the Ga sublattice have strength stronger than those on the As sublattice in the conduction LLs, and the increase of $p_x$- and $p_y$-orbital strength with the decrease of $s$-orbital strength take place as $n_1^{\uparrow}$ and $n_1^{\downarrow}$ grow [comparison between Figs.~\ref{fig:Geo_LL_WF_monoGaAs}(c) and (d) and Figs.~\ref{fig:2D_Geometry_BS_WF}(d)--(g)].
Instead of the Ga sublattice, the dominated orbitals on the As sublattice in the valence LLs have larger strength.
The $p_x$- and $p_y$-orbitals on a specific sublattice are of the same strength in the $n_2^{\downarrow} = 0$ and $n_3^{\uparrow} = 0$ valence LLs [Figs.~\ref{fig:Geo_LL_WF_monoGaAs}(g) and (j)], reflecting the fact that the zero-field $p_x$- and $p_y$-orbitals have equivalent strength at the $\Gamma$ point [Figs.~\ref{fig:2D_Geometry_BS_WF}(h)--(o)].
As the subband index increases, the $n_2^{\uparrow\downarrow}$ ($n_3^{\uparrow\downarrow}$) valence LLs become $p_y$-orbital- ($p_x$-orbital-) dominated, which resembles the $\mathbf{k}$-dependance of dominated orbitals near the $\Gamma$ point.
The aforementioned LL node regularities and energy-dependent orbital variation give a fundamental understanding for further researches in optical and transport properties, such as magneto-optical absorption selection rules including major/minor optical transitions and the possible/forbidden transport channels.

\begin{figure}
  \includegraphics[]{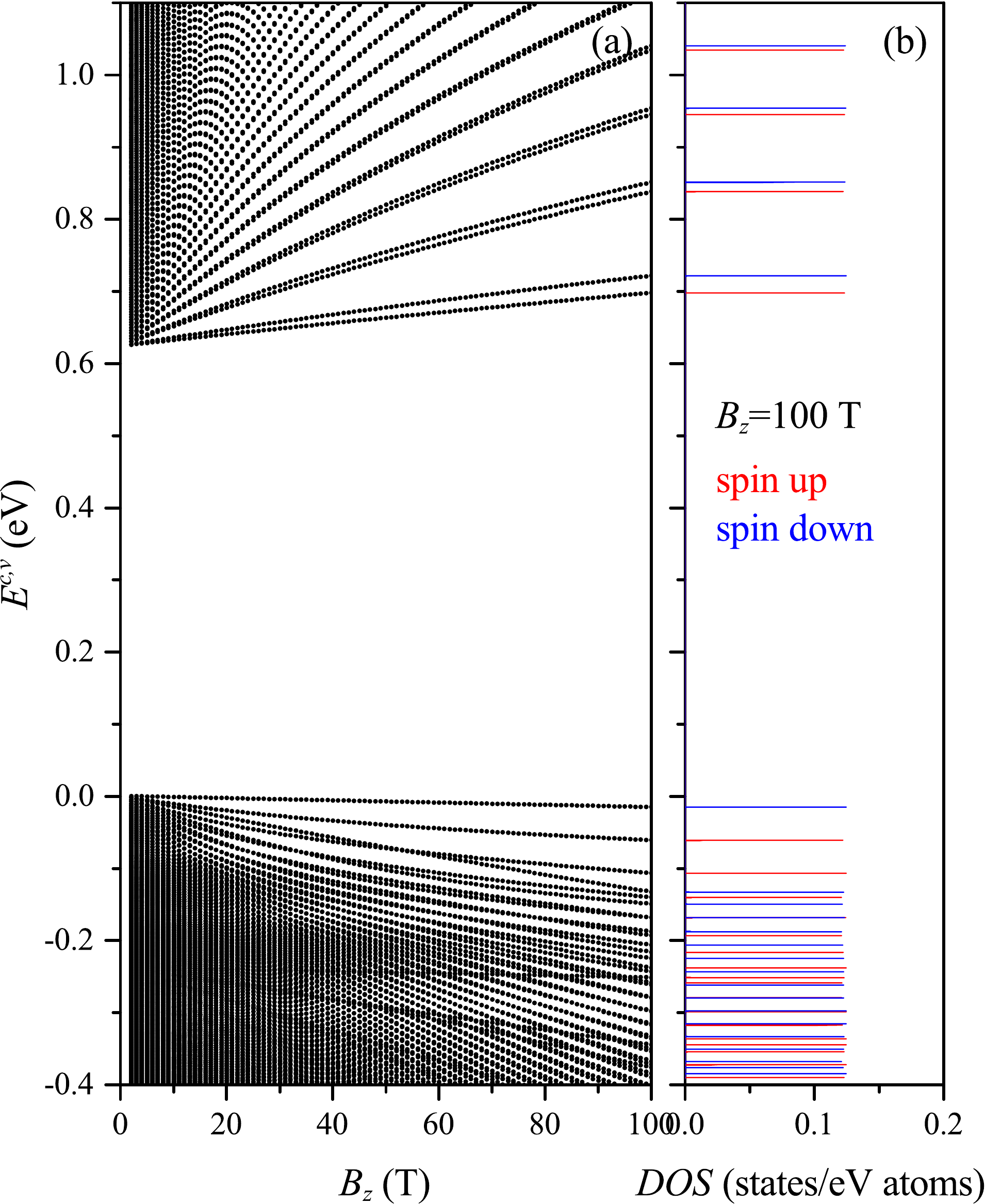}\\
  \caption{
(Color online)    
(a) Magnetic-field-dependent LL energy spectra of various groups.
(b) Spin-polarized DOS at $B_z = 100$ T.
}
  \label{fig:B_dependent_LL_monoGaAs}
\end{figure}

The low-lying LLs exhibit a spin-polarized $B_z$-dependent energy spectrum, as clearly shown in Fig.~\ref{fig:B_dependent_LL_monoGaAs}(a).
All LLs have a monotonic variation to the magnetic field, which reflects the monotonic band structure near the $\Gamma$ point at zero field.
In each spin-up (or spin-down) LL subgroup, the energy spacing between LLs is enlarged when the magnetic field increases.
%The $B_z$ dependence of the energy spacing is approximately linear owing to the parabolic subband at $B_z = 0$.
The $B_z$ dependence of the energy spacing is approximately linear owing to the parabolic energy dispersion at $B_z = 0$.
%However, the Bz dependence of the latter is approximately linear. due to parabolic subband
Between the $n$th spin-up and spin-down LLs of the same group ($n_i^{\uparrow} = n_i^{\downarrow} = n; i \in \{1,2,3\}$), their energy spacing grows with the increment of $B_z$, arising from the enhanced SOC by the more localized LL wave functions.
%With the increment of $B_z$, the energy spacing between the nc Gamma down and nc Gamma up LLs grows, arising from the enhanced Vsoc by the more localized LL wave functions.
For instance, the spacing is $24$ meV between $n_1^{\uparrow} = 0$ and $n_1^{\downarrow} = 0$ LLs at $B_z = 100$ T (comparable to the room temperature thermal energy). %(!!!!!!!!!other spin up and down energy spacing   > 24 meV is better)
At small magnetic field ($B_z \rightarrow 0$), the energy spacing between the lowest conduction LL and the highest valence LL closes to the zero-field energy gap.
For an increasing magnetic field, the gap gradually increases due to the rising of $n_1^{\uparrow} = 0$ LL state energy and the falling of $n_2^{\downarrow} = 0$ LL state energy.
%The main features of the LL spectrum are reflect on the DOS [Fig.~\ref{fig:B_dependent_LL_monoGaAs}(b)].
The DOS, defined as $\sum_{\mathbf{k}} \sum_{n_i^{\uparrow \downarrow}; i \in \{1,2,3\}} \delta [\omega - E^{c,v}(\mathbf{k}, n_i^{\uparrow \downarrow})]$, directly reflects the main features of the LL energy spectra as depicted in Fig.~\ref{fig:B_dependent_LL_monoGaAs}(b)~\cite{Phys.Chem.Chem.Phys.18(2016)7573H.C.Chung, Sci.Rep.5(2015)9423P.Y.Lo, Carbon109(2016)883H.C.Chung}.
%~\cite{Sci.Rep.5(2015)9423P.Y.Lo}.
%Three groups of LL peaks appear in the delta-function-like symmetric structure, in which their heights are proportional to state degeneracy.
%Three groups of LL peaks appear in the delta-function-like symmetric structure.
Three groups of delta-function-like symmetric peaks respectively appear from $\sim 0.62$ eV, $0$ eV, and $-0.24$ eV.
The delta-function-like symmetric peaks have two-side-divergent structure at the peak frequency $\omega$, i.e., the delta-function-like peak is symmetric about the axis of $\omega$.
Their peak heights are the same, indicating the identical degeneracy of LLs.
In each spin-polarized LL subgroup, the peak spacing is shrunk for a larger subband index.
The above-mentioned characteristics of LL peaks, including peak structure, height, and spacing, can be verified through the experimental measurements using STS~\cite{Phys.Rev.Lett.102(2009)176804G.Li, Science324(2009)924D.L.Miller, Nature467(2010)185Y.J.Song, Phys.Rev.Lett.106(2011)126802A.Luican, Phys.Rev.B92(2015)165420W.X.Wang}.
Furthermore, it is predicted that the optical absorption peaks are contributed by transitions between high-intensity-DOS LLs.

%----------------------------------------

%Ez-dependent LL

%----------------------------------------
%QUESTION: Does Ez cause an apparent of the band/spin splitting?

\begin{figure*}
  \includegraphics[]{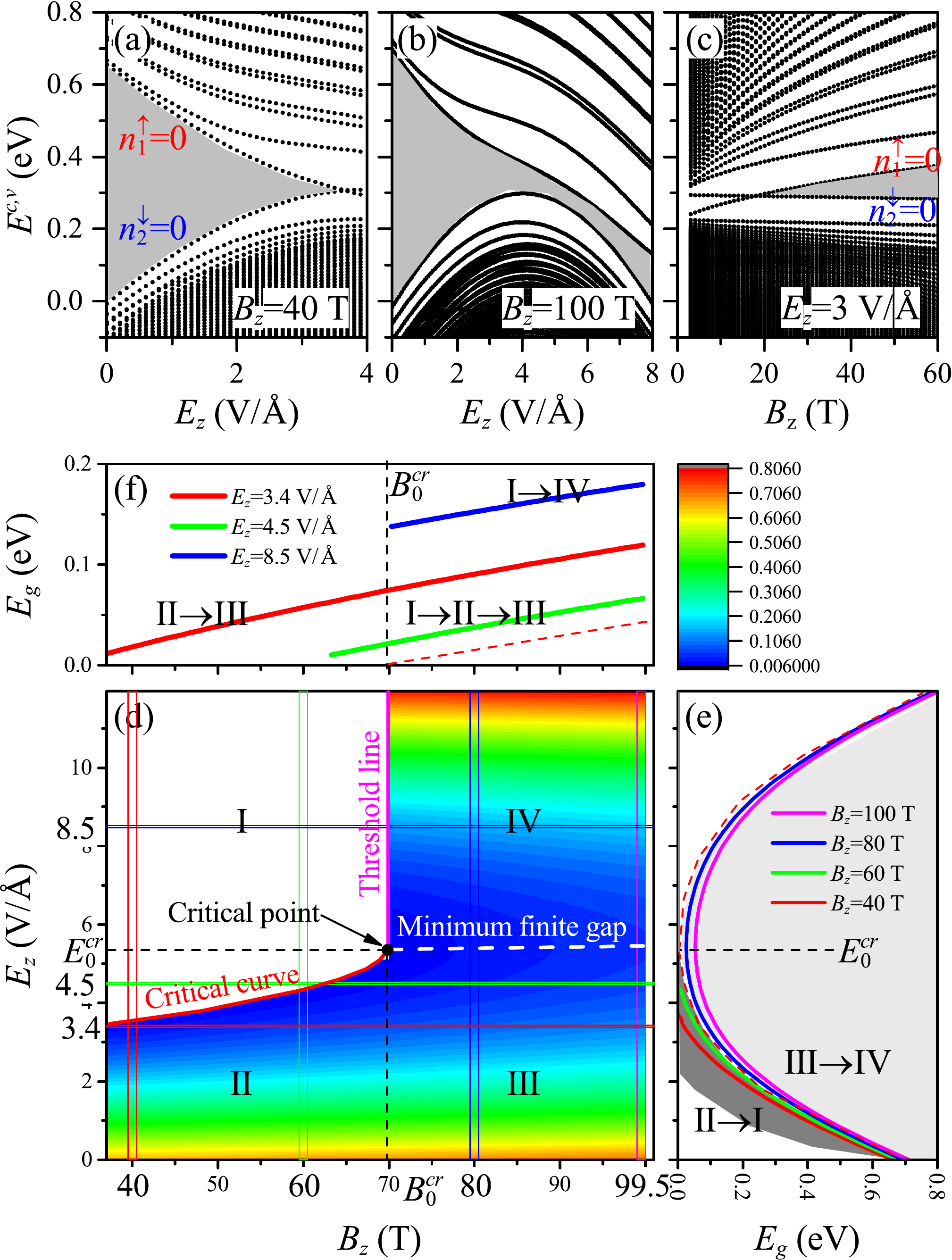}\\
  \caption{
(Color online) 
(a), (b) Gate-voltage-dependent LL energy spectra at $B_z = 40$ and $100$ T.
(c) Magnetic-field-dependent LL energy spectrum at $E_z = 3$ V/{\AA}.
(d)--(f) Dependence of energy gap on magnetic field and gate voltage.
}
  \label{fig:Eg_LL_map_monoGaAs}
\end{figure*}

The magneto-electronic properties of monolayer GaAs with buckled structure, which are much different from those of monolayer graphene with planer structure, can be diversified by a perpendicular electric field, $E_z$.
An electric potential difference $V_z = E_z l_z$ between the planes of Ga and As sublattice can cause monotonous/nonmonotonous dispersion relations, crossing LL spectra, enhancement of spin splitting, and modulation of energy gap.
For a small magnetic field [Fig.~\ref{fig:Eg_LL_map_monoGaAs}(a)], the $n_1^{\uparrow\downarrow}$ and $n_2^{\uparrow\downarrow}$/$n_3^{\uparrow\downarrow}$ LLs respectively exhibit monotonous decrease and increase as the electric field grows.
An intergroup LL crossing takes place between $n_1^{\uparrow} = 0$ and $n_2^{\downarrow} = 0$ LLs at the critical electric field ($E_z^{cr} = 3.6$ V/{\AA}), and their energy gap shrinks to zero (gray zone).
Meanwhile, the spin splitting is enhanced, where the energy spacing between the $n_1^{\uparrow} = 0$ and $n_1^{\downarrow} = 0$ LLs is about 100 meV (larger than the room temperature thermal energy).
For a large magnetic field [Fig.~\ref{fig:Eg_LL_map_monoGaAs}(b)], the $n_1^{\uparrow\downarrow}$ LLs present a monotonous decrease, while the $n_2^{\uparrow\downarrow}$/$n_3^{\uparrow\downarrow}$ LLs vary nonmonotonously with turning points (at $E_z \sim 4$ V/{\AA}).
%There is no intergroup crossing between $n_1^{\uparrow\downarrow}$ and $n_2^{\uparrow\downarrow}$ LLs, and the energy gap gradually shrinks and reaches a minimum none-zero value near the turning point.
The energy gap gradually shrinks and reaches a minimum finite value without the intergroup crossing between $n_1^{\uparrow\downarrow}$ and $n_2^{\uparrow\downarrow}$ LLs.
Magnetic fields can shift the gap of a top-gated monolayer GaAs.
As the magnetic field grows over the critical strength [$B_z^{cr} \sim 18$ T in Fig.~\ref{fig:Eg_LL_map_monoGaAs}(c)], the gap is opened and then gradually increases.

The gap modulation owing to the the competition between magnetic and electric fields is presented in detail by the color map, served as the $E_z$-$B_z$ phase diagram [Fig.~\ref{fig:Eg_LL_map_monoGaAs}(d)].
%[4 phase boundaries, 4 regions, 1 critical point, 3 phases]
There are four regions (I, II, III, and IV) separated by four boundaries (the critical curve, threshold line, threshold extension line, and finite minimum gap curve).
The red critical curve with an upward trend indicates the critical electric and magnetic fields ($E_z^{cr}$ and $B_z^{cr}$), where the system is a gapless semiconductor of $E_g^{SO} = 0$.
The vertical threshold line at the maximum critical magnetic field, $B_0^{cr}$, figures out a drastic change of the gap between finite value and zero.
%(, in which... 0 jump to finite value)
The red critical curve and the vertical threshold line intersect at the critical point, ($B_0^{cr}, E_0^{cr}$) = (70 T, 5.3 V/{\AA}), indicating the maximum critical magnetic and electric fields.
The curve on the right side of the critical point marks finite minimum gap during the competition.
%On the right side of the critical point is the curve marking finite minimum gap during the competition.
The semimetallic and semiconducting phases are respectively in regions I and (II, III, IV).
Also, the electric field is in competition with the magnetic field, inducing complex responses to the gate voltage in various regions.
For a specific magnetic field, the overlap between conduction and valence LLs in the region-I semimetal increases as the electric field grows.
The gap of the region-II, region-III, and region-IV semiconductors respectively, shrinks to zero, decreases to a finite minimum value, and increases from a finite minimum value for a growing $E_z$.
The diverse phase transitions occur from region I to regions (II, III, IV), e.g., a phase transition from semiconductor, gapless semiconductor to semimetal for regions II $\rightarrow$ I, and a transition from semimetal to semiconductor for I $\rightarrow$ IV.
%The color map gives the detailed information about the gap modulation via external fields.
Abundant gap modulation can be achieved by controlling the external fields.
The gate-voltage-controlled gap presents two types of modulation, associated to different region-to-region variations. [Fig.~\ref{fig:Eg_LL_map_monoGaAs}(e)].
At $B_z < B_0^{cr}$ (red and green curves), the $E_z$-dependent gap gradually shrinks to zero, where a large $B_z$ corresponds to a large $E_z^{cr}$ (cutoff points of the curves).
At $B_z > B_0^{cr}$ (blue and magenta curves), the gap reduces to a finite value and then increases, where the red dashed curve shows the lower limit of such modulation and the finite minimum value grows with the increasing $B_z$.
The ranges of gap modulation from regions II $\rightarrow$ I and III $\rightarrow$ IV are illustrated by the dark and light gray zones, respectively.
The magnetic-field-controlled gap possesses three types of modulation [Fig.~\ref{fig:Eg_LL_map_monoGaAs}(f)].
At $E_z < E_{min}^{cr}$ (red curve), the $B_z$-dependent gap gradually increases (a variation from regions II $\rightarrow$ III), where $E_{min}^{cr}$ is the critical electric field at $B_z \rightarrow 0$.
At $E_{min}^{cr} < E_z < E_0^{cr}$ (green curve), the gap is opened at $E_z^{cr}$ and increases gradually from zero (from regions I $\rightarrow$ II $\rightarrow$ III).
At $E_z > E_0^{cr}$ (blue curve), the opened gap gradually increases from a finite value at $B_0^{cr}$ (from regions I $\rightarrow$ IV).
The aforementioned external-field-controlled gap modulation and phase transitions are helpful in developing the top-gated electronic/optical devices and enable potential applications in phase-change electronic devices~\cite{Nat.Commun.7(2016)10671Y.Li}.

%EF NEED adjusted
%The longitudinal resistance Rxx is represented as a function of the external magnetic field B and for different gate voltages Vg.

%-----------------------------------------
%
%EXP (STS)
%What is STS
%Exp evidence of sqrt(B) and (B)
%Predicted properties can be identified
%
%-----------------------------------------

The main characteristics and the $B_z$-$E_z$-competition-induced modulation of symmetric Landau peaks in the DOS can be verified by STS.
It is an extension of scanning tunneling microscopy (STM)~\cite{Phys.Rev.Lett.49(1982)57G.Binnig, Helv.Phys.Acta55(1982)726G.Binnig, Phys.Rev.Lett.50(1983)120G.Binnig} and provides detailed information about the DOS on a sample surface, such as silicon~\cite{Phys.Rev.Lett.56(1986)1972R.J.Hamers, Surf.Sci.181(1987)295R.M.Feenstra, Phys.Rev.B48(1993)17892M.N.Piancastelli} and CNTs~\cite{Nature391(1998)59J.W.G.Wildoer, Phys.Rev.Lett.82(1999)1225P.Kim, Science283(1999)52L.C.Venema, Phys.Rev.B62(2000)5238L.C.Venema}.
The tunneling differential conductance ($dI/dV$), proportional to the DOS~\cite{Phys.Rev.B31(1985)805J.Tersoff}, directly reveals the main characteristics, i.e., the structures, positions, and intensities of the peaks.
Part of theoretical predictions on the LL energy spectra of few-layered graphene are verified, such as the $\sqrt{B_z}$-dependent LLs in monolayer graphene~\cite{Phys.Rev.Lett.102(2009)176804G.Li, Science324(2009)924D.L.Miller, Nature467(2010)185Y.J.Song, Phys.Rev.Lett.106(2011)126802A.Luican, Phys.Rev.B92(2015)165420W.X.Wang}, the linear $B_z$-dependent LLs in AB-stacked bilayer graphene~\cite{Phys.Rev.Lett.100(2008)087403E.A.Henriksen, Nat.Phys.7(2011)649G.M.Rutter, Phys.Rev.B93(2016)125422L.J.Yin}, the concurrence of square-root and linear $B_z$-dependent LLs in graphene of trilayer ABA stacking~\cite{Phys.Rev.B91(2015)115405L.J.Yin}.
The predicted magneto-electronic properties of the monolayer GaAs, including three groups of LLs with linear $B_z$-dependence, the external-field-controlled gap modulation and the SOC-induced spin splitting, could be further identified.
%The STS examinations can provide the critical informations in the lattice symmetry, stacking configuration, SOC, and single- or multiorbital hybridization.

The aforementioned main features of wave functions can be confirmed by spectroscopic-imaging STM~\cite{Science262(1993)218M.F.Crommie, Nature427(2004)328S.Ilani, Science327(2010)665A.Richardella}, which can resolve charge distributions from the local DOS and is an appropriate experimental technique for identifying standing waves and Landau wave functions on the surfaces of various condensed-matter systems.
Standing waves have been directly observed at the surface steps of Au(111) and Cu(111)~\cite{Nature363(1993)524M.F.Crommie, Science262(1993)218M.F.Crommie}, as well as finite-length metallic CNT~\cite{Science283(1999)52L.C.Venema, Phys.Rev.Lett.93(2004)166403J.Lee}.
%Furthermore, theoretical predictions of the standing waves in a finite-length metallic CNT~\cite{Phys.Rev.Lett.82(1999)3520A.Rubio, Phys.Rev.B65(2002)245418J.Jiang} have also been verified~\cite{Science283(1999)52L.C.Venema, Phys.Rev.Lett.93(2004)166403J.Lee}.
Also, the spatial mapping of the electronic states in the troughs between self-organized Pt nanowires on Ge(001) is presented~\cite{NanoLett.6(2006)1439A.vanHouselt, J.Phys.-Condens.Matter25(2013)014014R.Heimbuch}.
Recently, Landau orbits without nodes have been observed~\cite{Phys.Rev.Lett.101(2008)256802K.Hashimoto, Nat.Phys.6(2010)811D.L.Miller}, and subsequently, observations of the concentric-ring-like nodal structures have also been obtained~\cite{Phys.Rev.Lett.109(2012)116805K.Hashimoto, Nat.Phys.10(2014)815Y.S.Fu}.
In monolayer GaAs, the predicted orbital domination for various groups of LLs and the relative strength of various orbitals (or different sublattices) for a specific LL could be examined through spectroscopic-imaging STM measurements on nodal structures.

Monolayer graphene and GaAs have much different essential properties and responses to external fields based on the orbital domination, SOC, and geometric structure.
The low-energy electronic structure of planar graphene exhibits a pair of single-orbital-dominated ($p_z$) conduction/valence spin-degenerate subbands touching at the K point, resulting in a zero-gap.
In contrast to the gapless semiconducting graphene, monolayer buckled GaAs possesses a direct energy gap at the $\Gamma$ point among SOC-induced multi-orbital-dominated ($s$, $p_x$, and $p_y$) spin-polarized subbands, whose spin-splitting energies are $\mathbf{k}$-dependent.
Distinct features are revealed in magnetic quantizations, such as the magnetic field dependence of LLs, localization centers of Landau wave functions, and quantum mode regularities.
The LLs reflect the main features of zero-field energy dispersions, exhibiting $\sqrt{B_z}$-dependent spin-degenerate LLs and linear $B_z$-dependent spin-polarized LLs in monolayer graphene and GaAs, respectively.
The localization centers of the spin-degenerate states (spin-polarized states) are at $1/6$, $2/6$, $4/6$, and $5/6$ ($0$ and $1/2$) positions of the enlarged unit cell.
For a specific LL of graphene (GaAs), the major node numbers of subenvelope functions in different sublattices differ by one (are identical).
Electric fields further cause an on-site energy difference between two distinct sublattices of GaAs with buckled structure, leading to the gap modulation, phase transition, and enhancement of spin splitting.
The aforementioned differences clearly illustrate that electronic properties are diversified by the geometric structures, orbital hybridizations, spin configurations, as well as electric and magnetic fields.

%%-----------------------------------------
%Graphene              GaAs
%monoelemental 2D hexagonal               binary compounds of group III-V ele-
%ments
%single-orbital-dominated     multi-orbital-dominated
%
%pz orbital                s px py
%K point                   Gamma point
%low-energy --  pair c/v subbands   three subbands
%without SOC          SOC
%no spin splitting       with k-dependent spin splitting
%gapless semiconductor         gap
%
%unit cell 2 atoms
%
%Magnetic quantization
%field dependence    sqrt(B) LLs         linear B  spin-polarized LLs
%localization center 1/6 2/6 4/6 5/6     0 1/2   positions of the enlarged unit cell.
%quantum modes differ by one       identical quantum mode
%
%B+E
%planar structure    buckled structure   different on-site energies
%
%      enhancement of spin splitting
%gap modulation and phase transition
%%-----------------------------------------

\section{Conclusion}

We develop the generalized tight-binding model to study the essential properties of monolayer GaAs.
Many critical factors, including the buckled structure, multi-orbital hybridization, SOC, electric field, and magnetic field, are considered in the calculation simultaneously.
This system in contrast to graphene is predicted to have rich and unique magnetic quantizations and phase transitions.
%, clearly illustrating the important differences between GaAs and graphene.
The developed generalized tight-bind model provides a theoretical framework for investigating the competitions among various critical factors and affords systematic studies from multi-dimensional materials to hybrid systems.
Theories with both single-particle and many-body schemes can also be combined to comprehend the essential physical properties, e.g., frequency-dependent and static Kubo formulas for exploring the optical absorption spectra~\cite{Opt.Express19(2011)23350H.C.Chung, Phys.Chem.Chem.Phys.17(2015)26008C.Y.Lin, Phys.Chem.Chem.Phys.18(2016)7573H.C.Chung, Carbon109(2016)883H.C.Chung} and quantum Hall effect [arXiv:1704.01313], respectively.

%This model could directly combine with the other theories to comprehend the essential physical properties. For example, the combination with the frequency-dependent and static Kubo formulas is, respectively, useful in exploring the magneto-optical absorption spectra (Lin's papers) and quantum Hall effect (Do & Lin, 2017 e-print).

%-----------------------------------------

Band structures and LLs of monolayer GaAs are very sensitive to the buckled structure, multi-orbital hybridizations, spin-orbital interactions, and external fields.
Three groups of SOC-induced spin-polarized subbands ($n_1^{\uparrow \downarrow}$, $n_2^{\uparrow \downarrow}$, $n_3^{\uparrow \downarrow}$) initiated from the $\Gamma$ point exhibit monotonous energy dispersions and strong $\mathbf{k}$-dependent spin splitting.
There are a direct band gap ($E_g^{SO} = 0.623$ eV) between $n_1^{\uparrow \downarrow}$ and $n_2^{\uparrow \downarrow}$ subbands as well as a SOC-induced energy splitting ($\Delta_{SO} = 0.237$ eV) between $n_2^{\uparrow \downarrow}$ and $n_3^{\uparrow \downarrow}$ subbands at the $\Gamma$ point.
The whole-range state probabilities presenting the detailed orbital variations on different sublattices of various subbands are illustrated, showing that the conduction $n_1^{\uparrow \downarrow}$ subbands are $s$-orbital-dominated with larger state probabilities on the Ga sublattice; the valence $n_2^{\uparrow \downarrow}$ ($n_3^{\uparrow \downarrow}$) subbands are $p_y$-orbital- ($p_x$-orbital-) dominated with larger probabilities on the As sublattice.
Magnetic quantization induces three groups of spin-polarized LLs with initial energies respectively near $0.62$ eV, $0$ eV, and $0.24$ eV and a gap of size $\sim E_g^{SO}$ between the lowest conduction and highest valence LLs, reflecting the energies of zero-field electronic states at the $\Gamma$ point.
Each LL is doubly degenerate based on one $\Gamma$ valley and mirror symmetry.
The LL energy spacing for any particular spin-polarized subgroup gradually shrinks as the state energy grows.
The state probabilities of subenvelope functions are well-behaved in their spatial distributions, possessing oscillation patterns with regular nodes at the localization centers, similar to those of harmonic oscillators.
The doubly degenerate spin-polarized LL states at $(k_x, k_y) = (0, 0)$ are localized at the $0$ and $1/2$ positions of the enlarged unit cell, respectively.
In each LL, the node numbers of various orbital subenvelope functions on the Ga and As sublattices are identical, and the $s$-orbital node number differs the $p_x$-orbital ($p_y$-orbital) node number by one.
The orbital domination and the complex variation about the domiated/minor orbitals feature the average of accumulated neighboring zero-field electronic states, i.e., the energy-dependent relative orbital strength roughly corresponds to the $\mathbf{k}$-dependent relative orbital strength at $B_z = 0$ owing to the monotonous zero-field band structure near the Fermi level.
%monotonous zero-field low-lying subbands.
These predicted characteristics of state probabilities could be examined through spectroscopic-imaging STM measurements on nodal structures.
%In monolayer GaAs, the predicted orbital domination for various groups of LLs and the relative strength of various orbitals (or different sublattices) for a specific LL could be examined through spectroscopic-imaging STM measurements on nodal structures.

The linear-$B_z$ dependence of LL energies is revealed owing to the low-lying parabolic energy dispersions.
For an increasing magnetic field, the gap is enlarged and the spin splitting is enhanced gradually.
%the gap is enlarged gradually with the enhancement of the spin splitting.
There are three group of spin-polarized LL peaks in the DOS, directly reflects the main features of the LL energy spectra.
The delta-function-like symmetric peak structure, initial frequencies for each group of LL peaks, degeneracy-related peak height, and shrunk frequency spacings for peaks of larger indices could be identified by the STS measurements.

The electric field, leading to an electric potential difference in the buckled structure, causes monotonous/nonmonotonous energy dispersions, LL crossing, enhancement of spin splitting, and gap modulation.
For a small magnetic field $B_z < B_0^{cr}$, the intergroup LL crossing occurs between the conduction and valence LLs at the critical electric field, accompanied by the gap shrinkage and close; for a large magnetic field $B_z > B_0^{cr}$, the gap remains finite without the occurrence of intergroup LL crossing near the Fermi level.
It should be noted that the spin splitting is enhanced with an energy spacing larger than the room temperature thermal energy.
The complex gap modulations and phase transitions based on the competition between magnetic and electric fields are investigated.
The $E_z$-$B_z$ phase diagram illustrates the complex phase transitions between four characteristic regions.
The $E_z$- ($B_z$-) controlled gap presents two (three) types of modulation, associated to different region-to-region variations.
The field-controlled gap modulations and phase transitions are helpful in developing the top-gated electronic/optical devices and phase-change electronic devices.
%The field-controlled gap modulations and phase transitions are helpful in developing the top-gated electronic/optical devices [] and phase-change electronic devices [].

Monolayer GaAs, being 2D materials beyond graphene, is much different from graphene on the essential properties and responses to external fields owing to the orbital domination, SOC, and geometric structure.
Distinct magnetic quantizations are revealed, such as the magnetic field dependence of LLs, localization centers of LL wave functions, and quantum mode regularities.
Electric fields, leading to an on-site energy difference in GaAs with buckled structure, further induce the gap modulation, phase transition, and enhancement of spin splitting.
The predicted magneto-electronic properties of the monolayer GaAs, including three groups of spin-polarized LLs with linear $B_z$-dependence, the external-field-controlled gap modulation/phase transition and the SOC-induced spin splitting, could be further experimentally identified by STS.
Additionally, this work can be treated as a model study for comprehending magnetic quantizations of other group III-V 2D materials.

\begin{acknowledgments}
\begin{CJK}{UTF8}{bsmi}
We would like to thank all the contributors to this article for their valuable discussions and recommendations, especially Geng-Ming Hu, Matisse Wei-Yuan Tu, Ping-Yuan Lo, and Yu-Ming Wang.
The authors thank Pei-Ju Chien for English discussions and corrections.
One of us (Hsien-Ching Chung) thanks Ming-Hui Chung, Su-Ming Chen, Lien-Kuei Chien, Mi-Lee Kao, and Fu-Long Chen for financial support.
%One of us (Hsien-Ching Chung) thanks Ming-Hui Chung (鍾明輝), Su-Ming Chen (陳素敏), Lien-Kuei Chien (簡聯貴), Mi-Lee Kao (高蜜李), Fu-Long Chen (陳富隆) for financial support.
This work was supported in part by the Gin Wen Town Printing Company, Taichung, Taiwan.
This work was supported in part by the Ministry of Science and Technology of Taiwan under grant numbers MOST 105-2811-M-017-003, MOST 105-2112-M-017-002-MY2, and NSC 102-2112-M-006-007-MY3.
\end{CJK}
\end{acknowledgments}

%胡耿銘
%涂維元 Matisse Wei-Yuan Tu
%羅炳蒝 Ping-Yuan Lo
%王欲銘 Yu-Ming Wang

\bibliography{Reference_monoGaAs}
\bibliographystyle{apsrev4-1}

\end{document}